\documentclass[twocolumn,amsmath,amssymb,aps,pra,floatfix]{revtex4}
\usepackage{bm}
\usepackage{amsfonts}
\usepackage{amsmath}
\usepackage{graphicx}
\usepackage{epstopdf}
\begin{document}
\title{Guiding neutral particles endowed with a magnetic moment by an electromagnetic wave carrying orbital angular momentum. II. Quantum mechanics}
\author{Tomasz Rado\.zycki}
\email{torado@fuw.edu.pl}
\affiliation{Faculty of Mathematics and Natural Sciences, College of Sciences, Cardinal Stefan Wyszy\'nski University, W\'oycickiego 1/3, 01-938 Warsaw, Poland} 
\begin{abstract}
The quantum mechanical states of the neutral particle endowed with a magnetic moment in the combination of electromagnetic vortex field together with the constant magnetic field are dealt with. It is shown that this system of fields is capable to capture the particle in the perpendicular direction and guide it along the propagating wave. The quantum evolution is subject to tunneling processes, which can destroy the delicate trapping mechanism. The probability of these processes show that it should be in principle possible to catch and guide the particle for the time of several microseconds. This time can be lengthened by the appropriate adjustments of the external magnetic field.
\end{abstract}
\maketitle

\section{Introduction}

Guiding neutral particles, especially atoms or molecules, by light beams has attracted much attention for the last 20 years. It refers both to experimental~\cite{schi,kupp,rho1,glo,rho2} and theoretical investigations~\cite{shan,mark,renn1,renn2,yin,arlt1,arlt2}. A light beam used for transporting particles may be for instance composed of evanescent modes propagating along a hollow fiber or live in free space and form a kind of an optical vortex. The mechanism of confining atoms and molecules, which are neutral objects and do not interact directly with electric field of the wave, is based on the Stark effect. It leads to the emergence of a certain binding potential in the direction perpendicular to that of the wave propagation and allows for guiding atoms along the light beam. Another arrangement exploits the rotating magnetic field for binding both charged and neutral particles~\cite{mielnik,fer,QL,shapiro}.

In our previous work~\cite{ibbtr2} we proposed another subtle mechanism which allows to guide or even trap neutral particles due to the interaction of their magnetic moments with the magnetic field of the vortex. This mechanism does nor refer to the particle internal structure and, therefore, can in principle be applied equally well to composed objects like atoms or positronium and elementary ones like neutrons (their quark structure is irrelevant here). The binding effect of the vortex field can be reinforced by the presence of the external uniform magnetic field pointing along the wave-propagation direction. The {\em classical} equations of motion of the particle were solved in the quoted work and the obtained trajectories explicitly showed that this configuration of fields is capable to trap or guide particles.

The true theory governing processes occurring in micro-world has, however, the quantum nature. Therefore in the present work we address this problem within quantum mechanics. The Schr\"odinger-Pauli equation for a neutral particle endowed with magnetic moment $\mu$ has the form:
\begin{align}\label{sp}
i\hbar\frac{\partial}{\partial t}\Psi({\bm r},t) =\left(-\frac{\hbar^2\Delta}{2M}-g\bm{s}\!\cdot\!{\bm B}\right)\Psi({\bm r},t).
\end{align}
where $g$ denotes the gyromagnetic ratio (positive or negative depending on the orientation of the magnetic moment with respect to the spin angular momentum). For spin-$1/2$ particles, as the neutron for instance, the spin vector $\bm s$ is an operator expressed by Pauli matrices ${\bm s}=\hbar/2\,{\bm\sigma}$ and the wave-function $\Psi({\bm r},t)$ is two-component one. The same refers to a composed system as the hydrogen atom for instance (see Appendix). This Schr\"odinger equation can be simplified by the separation of time, $z$ and $\varphi$ variables (in cylindrical coordinates) and this is done in section~\ref{separ}. We are then left with two coupled ordinary differential equations in the radial variable for the upper and lower components of the spinor function.

In section~\ref{feq} further simplification is achieved by exploiting the diagonal form of the matrix potential. The obtained form constitutes the convenient starting point for the perturbative calculation. In the following section these results are used to analyze the motion of the particle. This analysis indicates the existence of bound states (in the perpendicular direction). These bound states can become unstable due to the possible tunneling effects characteristic for quantum physics. The possible tunneling channels are considered in subsections~\ref{tunnelb} and~\ref{tunnelc}.
Numerical estimations of the tunneling probabilities show that only the process consisting on flipping the spin direction can play an essential role, but the guiding time of hundreds of microseconds can in principle be achieved. This time may be lengthened by the appropriately adjusting the value of the external constant magnetic field. The similar effect can also be observed for the classical motion.

\section{Separation of variables}
\label{separ}

Consider the Bessel beam as given in~\cite{trojan} endowed with nonzero orbital angular momentum. It may be labeled with the value of $M-1$ (this $M$ appears only in this place and should not be confused with the particle mass used throughout the paper). We are particularly interested in the case $M=2$, dealt with in a series of our previous works~\cite{ibbprl,trojan,ibbtr1,ibbtr2}, in which case the field bears the vortex topological number equal to $1$. 

 This wave is accompanied the the external uniform magnetic field ($B_z$) oriented along the $z$ axis, the proper adjustment of which may help to create a stable trap. The total magnetic field in cylindrical coordinates $(\rho, \varphi, z)$ is then given by
\begin{align}\label{fieldfull}
{\bm B}({\bm r},t)=\left[\begin{array}{l}
-2B_\perp [a_+ \sin(\zeta_z-\varphi)J_1(k_\perp\rho)\\ \hspace{9ex} +a_- \sin(\zeta_z-3\varphi)J_3(k_\perp\rho)]\\
2B_\perp [a_+ \cos(\zeta_z-\varphi)J_1(k_\perp\rho)\\ \hspace{9ex} - a_- \cos(\zeta_z-3\varphi)J_3(k_\perp\rho)]\\
-4B_\perp \cos(\zeta_z-2\varphi)J_2(k_\perp\rho)+ B_z
\end{array}\right],
\end{align}
where  $J_i$ denotes the Bessel functions, $k=\sqrt{k_z^2+k_\perp^2}=\omega/c$ is the wave number, $B_\perp$ measures the strength of the vortex wave, $a_\pm=(\omega/c\pm k_z)/(2k_\perp)$ and $\zeta_z=\omega t-k_z z$.

In the paraxial approximation, where $k_\perp\rho\ll1$ and $k_\perp^2 z/k_z\ll 1$, we have:
\begin{align}
k_z\approx k,\;\;\;\; a_+\approx 1,\;\;\;\; a_-\approx 0,
\label{papp}
\end{align}
and the magnetic field may be written as~\cite{trojan}
\begin{align}\label{field}
{\bm B}({\bm r},t)=\left[\begin{array}{c}
B_\perp k(y\cos\,\zeta-x\sin\,\zeta)\\
B_\perp k(x\cos\,\zeta+y\sin\,\zeta)\\
B_z
\end{array}\right],
\end{align}
where $\zeta=\omega t-kz$. This last form can be now plugged into the Schr\"odinger equation. 

In order to prepare the separation of variables we shall make the following substitution:
\begin{align}\label{subst}
\Psi({\bm r},t)=e^{-i\zeta\sigma_z/2}\tilde{\Psi}({\bm r},t),
\end{align}
which eliminates the dependence of the Hamiltonian simultaneously on $t$ and $z$. Namely we obtain,
\begin{align}\label{sp1}
&i\hbar\frac{\partial}{\partial t}\tilde{\Psi}({\bm r},t) =\bigg[-\frac{\hbar^2\Delta_\perp}{2M}-\frac{\hbar^2(\partial_z+ik\sigma_z/2)^2}{2M}
\nonumber\\
&-\frac{\hbar \omega}{2}\,\sigma_z-\frac{g\hbar B_\perp k}{2}(x\sigma_y+y\sigma_x)-\frac{g\hbar B_z}{2}\,\sigma_z\bigg]\tilde{\Psi}({\bm r},t),
\end{align}
where $\Delta_\perp=\partial_x^2+\partial_y^2$.

Following~\cite{ibbtr2} and aimed at simplifying the equation~(\ref{sp1}), let us now introduce the dimensionless parameters:
\begin{subequations}\label{param}
\begin{align}
&\alpha=\frac{gB_\perp}{\omega}\sqrt{\frac{Mc^2}{\hbar \omega}},\;\;\;\;\; \beta=(1+\frac{g B_z}{\omega})\sqrt{\frac{Mc^2}{\hbar \omega}},\label{param1}\\ 
&\gamma=\frac{gB_\perp}{\omega},\;\;\;\;\;  \kappa_z=\frac{k_z}{k},\;\;\;\;\; {\cal E}_\perp=\frac{E_\perp}{\hbar\omega}\sqrt{\frac{Mc^2}{\hbar\omega}},\label{param2}
\end{align}
\end{subequations}
together with dimensionless time, space coordinates and momenta:
\begin{subequations}\label{dc}
\begin{align}
&{\bm \xi}=k{\bm r},\;\;\;\; \xi=k \rho,\label{dc1}\\
&\tau=\omega t\sqrt{\frac{\hbar \omega}{Mc^2}},\;\;\;\; {\bm \eta}=\frac{\bm p}{\sqrt{\hbar \omega M}},\label{dc2}
\end{align}
\end{subequations}
where $\rho=\sqrt{{\bm r}^2}$ and $\xi=\sqrt{{\bm\xi}^2}$. In this way we obtain in place of~(\ref{sp1}):
\begin{align}
i\frac{\partial}{\partial \tau}\tilde{\Psi}({\bm \xi},\tau) ={\cal H}\tilde{\Psi}({\bm \xi},\tau),
\label{sp2}
\end{align}
where the transformed Hamiltonian is given by 
\begin{eqnarray}
{\cal H}=&&\!\!\!\! -\frac{\gamma}{2\alpha}\,\Delta_{\xi_\perp}-\frac{\gamma}{2\alpha}\, (\partial_{\xi_z}+\frac{i}{2}\sigma_z)^2
-\frac{\beta}{2}\,\sigma_z\nonumber\\
&&\!\!\!\! -\frac{\alpha}{2}\,(\xi_x\sigma_y+\xi_y\sigma_x).
\label{hamilt}
\end{eqnarray}

Apart from~(\ref{hamilt}) there exist two other constants of motion:
\begin{subequations}\label{cm}
\begin{align}
H_1&={\bm \sigma}^2,\label{cm1}\\
H_2&=i(\xi_y\partial_{\xi_x}-\xi_x\partial_{\xi_y})-\frac{1}{2}\,\sigma_z,\label{cm2}
\end{align}
\end{subequations}
corresponding to the classical ones found in~\cite{ibbtr2}. The latter in polar coordinates reads: $H_2=-i\partial_\varphi -\sigma_z/2$. The fourth constant known from classical motion has already been exploited in~(\ref{subst}) while passing from $\Psi$ to $\tilde{\Psi}$. 

One should note that the classical change of variables: $({\bm x}, {\bm p})\mapsto ({\bm \xi}, {\bm \eta})$ was not canonical and, therefore, the quantum-mechanical commutator equals
\begin{align}
[\xi_m,\eta_n]=i\frac{\gamma}{\alpha}\, \delta_{mn},
\label{com}
\end{align}
and not simply $i\delta_{mn}$. It can be verified by a direct computation that one actually has
\begin{align}
[H_i,{\cal H}]=0, \;\;\;\;\mathrm{for}\; i=1,2.
\label{coh}
\end{align}

Exploiting (\ref{cm2}) we can substitute $\tilde{\Psi}$ in the form:
\begin{align}\label{sep}
\tilde{\Psi}({\bm \xi},\tau)=e^{-i{\cal E}_\perp\tau}e^{-i\gamma \kappa_z^2\tau/(2\alpha)}e^{i\kappa_z\xi_z}\left[\begin{array}{c}
e^{i(m+1)\varphi}f_+(\xi)\\
ie^{im\varphi}f_-(\xi)
\end{array}\right],
\end{align}
$m$ being an integer, and separate the remaining polar variables. This leads to two coupled ordinary differential equations:
\begin{subequations}
\label{sp3}
\begin{align}
-\frac{\gamma}{2\alpha}\bigg[f_+''&+\frac{1}{\xi}\, f_+'-\left(\frac{(m+1)^2}{\xi^2}+\kappa_z+\frac{1}{4}\right)f_+\bigg]\nonumber\\
&-\left({\cal E}_\perp+\frac{\beta}{2}\right)f_+=\frac{\alpha}{2}\,\xi f_-,\label{sp3a}\\
-\frac{\gamma}{2\alpha}\bigg[f_-''&+\frac{1}{\xi}\, f_-'-\left(\frac{m^2}{\xi^2}-\kappa_z+\frac{1}{4}\right)f_-\bigg]\nonumber\\
&-\left({\cal E}_\perp-\frac{\beta}{2}\right)f_-=\frac{\alpha}{2}\,\xi f_+,\label{sp3b}
\end{align}
\end{subequations}
where `prime' denotes now the derivative over $\xi$.

\section{The properties of the matrix potential}
\label{feq}

The first derivatives can be eliminated by plugging into~(\ref{sp3}) the functions $f_\pm(\xi)$ in the form:
\begin{align}
f_\pm(\xi)=\frac{F_\pm(\xi)}{\sqrt{\xi}},
\label{deff}
\end{align}
and we obtain the following equations for $F_\pm(\xi)$:
\begin{subequations}
\label{fs}
\begin{align}
-\frac{\gamma}{2\alpha}F''_++\frac{\gamma}{2\alpha}\bigg[&\frac{(m+3/2)(m+1/2)}{\xi^2}-\delta_+\bigg]F_+=\frac{\alpha}{2}\,\xi F_-,\label{fsa}\\
-\frac{\gamma}{2\alpha}F''_-+\frac{\gamma}{2\alpha}\bigg[&\frac{(m+1/2)(m-1/2)}{\xi^2}-\delta_-\bigg]F_-=\frac{\alpha}{2}\,\xi F_+,\label{fsb}
\end{align}
\end{subequations}
where 
\begin{align}
\delta_\pm=\frac{\alpha}{\gamma}(2{\cal E}_\perp \pm \beta )-\frac{1}{4}\mp\kappa_z.
\label{delta}
\end{align}
It may be easily verified that for the values of parameters considered in Section~\ref{pa} we have $\delta_\pm>0$.

\begin{figure*}[t]
\begin{center}
\includegraphics[width=0.8\textwidth,angle=0]{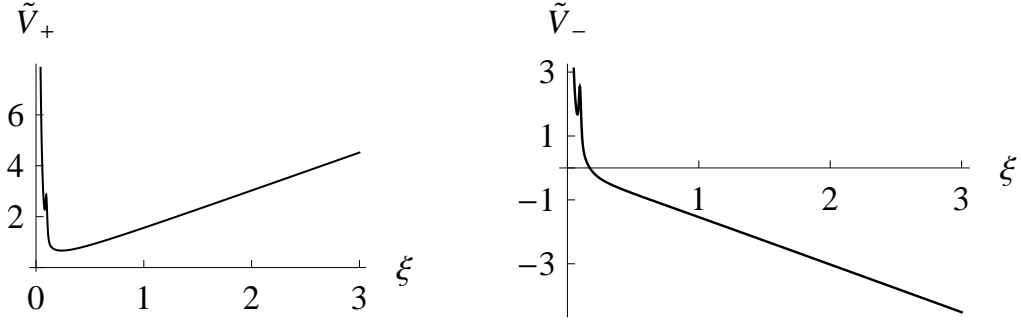}
\end{center}
\caption{Behavior of $\tilde{V}_\pm(\xi)$ for the values of parameters as in Fig.~5 of~\cite{ibbtr2}: $\alpha=3, \beta=0.8, \gamma=0.01$. Additionally $\kappa_z = 0.9, m = 2$.}
\label{vv1}
\end{figure*}

\begin{figure*}[t]
\begin{center}
\includegraphics[width=0.8\textwidth,angle=0]{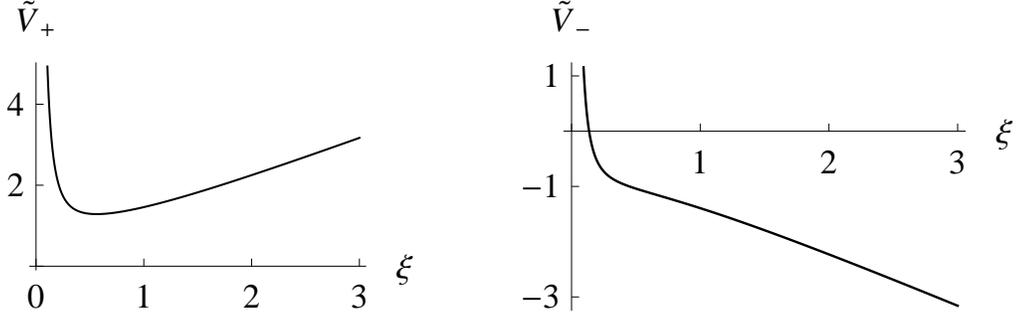}
\end{center}
\caption{Behavior of $\tilde{V}_\pm(\xi)$ for the values of parameters as in the last plot of Fig.~8 of~\cite{ibbtr2}: $\alpha=-2, \beta=-2, \gamma=-0.02$. Additionally $\kappa_z = 0.9, m = 2$.}
\label{vv2}
\end{figure*}

For the two-component function $F=[F_+,F_-]$ the set of equations~(\ref{fs}) may be given the form of a matrical stationary Schr\"odinger equation:
\begin{align}
-\frac{\gamma}{2\alpha}\partial_\xi ^2 F(\xi)+V(\xi)F(\xi)={\cal E}_\perp F(\xi),
\label{msch}
\end{align}
with the `potential'
\begin{widetext}
\begin{align}
V(\xi)=\left[\begin{array}{cc}\displaystyle \frac{\gamma(\kappa_z+1/4)}{2\alpha}-\frac{\beta}{2}+\frac{\gamma}{2\alpha}\, \frac{(m+3/2)(m+1/2)}{\xi^2} &\displaystyle  -\frac{\alpha\xi}{2} \\ \displaystyle -\frac{\alpha\xi}{2} & \displaystyle \frac{\gamma(-\kappa_z+1/4)}{2\alpha}+\frac{\beta}{2}+\frac{\gamma}{2\alpha}\, \frac{(m+1/2)(m-1/2)}{\xi^2} \end{array}\right],
\label{vpot}
\end{align}
\end{widetext}
and eigenenergy ${\cal E}_\perp$ to be determined. We will show below that this equation exhibits bound states, at least within perturbation theory.

The matrix $V$ is real, symmetric, and has the following eigenvalues
\begin{align}
V_\pm(\xi)=\frac{\gamma}{2\alpha}\left(\frac{(m+1/2)^2}{\xi^2}+\frac{1}{4}\right)\pm \Lambda(\xi), 
\label{ev}
\end{align}
where
\begin{align}
\Lambda(\xi)=\left[\Theta(\xi)^2+\frac{\alpha^2\xi^2}{4}\right]^{1/2},
\label{lambda}
\end{align}
and
\begin{align}
\Theta(\xi)= \frac{\gamma}{2\alpha}\left(\frac{m+1/2}{\xi^2}-\frac{\alpha\beta}{\gamma}+\kappa_z\right).
\label{theta}
\end{align}

The corresponding eigenvectors are
\begin{align}
\chi_\pm(\xi)=\left[\begin{array}{c}\displaystyle \Theta(\xi)\pm \Lambda(\xi)\\ \displaystyle -\frac{\alpha\xi}{2}  \end{array}\right]\sigma_\pm(\xi),
\label{funchi}
\end{align}
where $\sigma_\pm(\xi)$ are certain scalar functions defined below.
These eigenvectors $\chi_\pm(\xi)$ satisfy the conditions:
\begin{subequations}\label{orto}
\begin{align}
\chi_+(\xi)^T \chi_+(\xi)=\chi_-(\xi)^T \chi_-(\xi)=1,\label{orto1}\\
\chi_+(\xi)^T \chi_-(\xi)=\chi_-(\xi)^T \chi_+(\xi)=0,\label{orto0}
\end{align}
\end{subequations}
provided we choose 
\begin{align}
\sigma_\pm(\xi)=\left[2\Lambda(\xi)\left(\Lambda(\xi)\pm \Theta(\xi)\right)\right]^{-1/2}.
\label{sigma}
\end{align}

Let us now substitute into~(\ref{msch}) the wave-function in the form of $F(\xi)=U(\xi)\Phi(\xi)$, where $U$ is the $\xi$-dependent transformation matrix for the potential $V$:
\begin{align}
U^{-1}VU=V_D=\left[\begin{array}{cc} V_+ & 0\\ 0 & V_-\end{array}\right],
\label{tm}
\end{align}
and `$D$' stands for `diagonal'. The matrix $U$ is orthogonal and its columns constitute the eigenvectors $\chi_\pm$:
\begin{align}
U=\left[\begin{array}{cc} \chi_+,& \chi_-\end{array}\right],\;\;\;\mathrm{and}\;\;\; U^{-1}=U^T=\left[\begin{array}{c} \chi_+^T\\ \chi_-^T\end{array}\right].
\label{Umat}
\end{align}
The equation for the function $\Phi$ can be obtained in a straightforward way:
\begin{align}
-\frac{\gamma}{2\alpha}U^{-1}\partial_\xi^2 U\Phi +V_D\Phi={\cal E}_\perp\Phi,
\label{eqph}
\end{align}
and it is equivalent to
\begin{align}
-\frac{\gamma}{2\alpha}\,\partial_\xi^2\Phi+(V_D+W)\Phi(\xi)={\cal E}_\perp\Phi(\xi),
\label{eeq}
\end{align}
where the quantity
\begin{align}
W=\frac{\gamma}{2\alpha}\left(\partial^2_\xi-U^{-1}\partial^2_\xi U\right)
\label{vp}
\end{align}
may be treated as a perturbation for the diagonal potential $V_D$. It should be pointed out that the derivative in the expression $U^{-1}\partial^2_\xi U$ acts on both $U$ matrix and the wave-function in~(\ref{eeq}). 

If expressed through the eigenvectors~(\ref{funchi}) it has the formal form
\begin{eqnarray}
W&\!\!\!=&\!\!\!\frac{\gamma}{2\alpha}\left(\partial^2_\xi-\left[\begin{array}{c} \chi_+^T\\ \chi_-^T\end{array}\right]\partial^2_\xi \left[\begin{array}{cc} \chi_+,& \chi_-\end{array}\right]\right)\label{potw}\\
&\!\!\!=&\!\!\!-\frac{\gamma}{2\alpha}\left(\left[\begin{array}{cc} \chi_+^T\chi_+^{''} & \chi_+^T\chi_-^{''}\\ \chi_-^T\chi_+^{''} &  \chi_-^T\chi_-^{''}\end{array}\right] +2\left[\begin{array}{cc}\chi_+^T\chi_+' & \chi_+^T\chi_-'\\ \chi_-^T\chi_+' & \chi_+^T\chi_+'\end{array}\right] \partial_\xi\right),\nonumber
\end{eqnarray}

As it can be seen, this quantity contains both diagonal and off-diagonal elements and, therefore, it is more convenient to absorb all the diagonal terms (denoted below by $W_D$) into the unperturbed potential $V_D$, defining: 
\begin{align}
\tilde{V}_\pm(\xi)=V_\pm(\xi)-\frac{\gamma}{2\alpha}\chi_\pm^T(\xi)\chi_\pm^{''}(\xi).
\label{vt}
\end{align}
The behavior of $\tilde{V}_\pm(\xi)$ for various values of parameters is shown in Figures~\ref{vv1}-\ref{vv2}. 

The apparently diagonal elements remaining in the second matrix of~(\ref{potw}), i.e. the quantities containing $\chi_+^T\chi_+'$ and $\chi_-^T\chi_-'$, identically vanish as a consequence of normalization~(\ref{orto1}). The inclusion of $W_D$ into $\tilde{V}_D$ does not change the general form of $V_\pm$, since the additional terms disappear quickly as $\xi\rightarrow \infty$ and tend to constants for $\xi\rightarrow 0$. 

Now the perturbation responsible for the interaction between the two channels has become purely off-diagonal matrix $\tilde{W}$ of the form
\begin{align}
\tilde{W}=-\frac{\gamma}{2\alpha}\left(\left[\begin{array}{cc}0 & \chi_+^T\chi_-^{''}\\ \chi_-^T\chi_+^{''} & 0\end{array}\right] +2\left[\begin{array}{cc}0 & \chi_+^T\chi_-'\\ \chi_-^T\chi_+' & 0\end{array}\right] \partial_\xi\right).
\label{ww}
\end{align}
Upon omission of $\tilde{W}$ in the Schr\"odinger equation 
\begin{align}
-\frac{\gamma}{2\alpha}\,\partial_\xi^2\Phi+(\tilde{V}_D+\tilde{W})\Phi(\xi)={\cal E}_\perp\Phi(\xi)
\label{scht}
\end{align}
the two channels described by the lower and upper components of the wave-function $\Phi$ decouple from each other leading to the sector of well localized bound states living in the potential $\tilde{V}_+$, and the sector of scattering states governed by $\tilde{V}_-$. However in the practical situation of the vortex field, which is certainly limited in the direction perpendicular to the wave propagation, it would be unphysical, to expect these potentials to extend to infinity in the unmodified form. We will come back to that point in  the following section.

\section{Analysis of the motion}
\label{pa}

The Schr\"odinger equation (\ref{sp3}), (\ref{msch}) or (\ref{scht}) in its full complexity cannot be solved in an analytic way. Therefore, we consider the behavior of solutions separately for small and large radial distances and apply appropriate approximations.

\subsection{Small distances}
\label{sd}

Consider now the distances close to the vortex core. By `close' we mean $\xi\ll 1$, which roughly corresponds to $\rho\lesssim  \lambda$ (where $\lambda$ is the wavelength). In this region the equations~(\ref{sp3}) can directly be used. The term in square brackets of~(\ref{sp3a}) and analogously~(\ref{sp3b}) dominates over $\alpha\xi$ on the right-hand sides. We are then allowed to consider the simplified equations
\begin{subequations}
\label{simp}
\begin{align}
&f''_++\frac{1}{\xi}\, f'_+
+\bigg[\delta_+-\frac{(m+1)^2}{\xi^2}\bigg]f_+=0,\label{simpa}\\
&f''_-+\frac{1}{\xi}\, f'_-
+\bigg[\delta_--\frac{m^2}{\xi^2}\bigg]f_-=0.\label{simpb}
\end{align}
\end{subequations}

After the appropriate rescaling of the variable $\xi$, one can recognize in the above the ordinary and modified Bessel equations correspondingly. Therefore, their solutions can be immediately written out:
\begin{subequations}
\label{fpm1}
\begin{align}
f_+(\xi)&=C_1J_{m+1}(\sqrt{\delta_+}\, \xi)+C_2Y_{m+1}(\sqrt{\delta_+}\, \xi)\label{fpm1a}\\
f_-(\xi)&=D_1J_m(\sqrt{\delta_-}\, \xi)+D_2Y_m(\sqrt{\delta_-}\, \xi).\label{fpm1b}
\end{align}
\end{subequations}

To guarantee the correct behavior of the wave-function $\Psi$ at the origin, we must reject the term containing Neumann function $Y_{m+1}$ in $f_+$ and $Y_{m}$ in $f_-$, by putting $C_2=D_2=0$. It is then possible to choose the solution of~(\ref{sp1}), which is deprived of any ambiguities at the origin and is perfectly square-integrable as $\xi\rightarrow 0$. However, the small $\xi$ behavior does not rule on the trapping efficiency.

\subsection{Perturbative equations}
\label{peq}

For larger perpendicular directions the perturbative analysis based on Eq.~(\ref{scht}) is necessary. If $\tilde{W}=0$ there are two unperturbed one-dimensional Schr\"odinger equations constituting the starting point for the calculation, corresponding to the choice of either $\Phi_0=[u,0]$ or $\Phi_0=[0,v]$
and the potential $\tilde{V}_+$ or $\tilde{V}_-$ respectively:
\begin{subequations}\label{onedim}
\begin{align}
&-\frac{\gamma}{2\alpha}\partial_\xi^2 u(\xi)+\tilde{V}_+(\xi)u(\xi)={\cal E}_{\perp 0} u(\xi),
\label{onedimp}\\
&-\frac{\gamma}{2\alpha}\partial_\xi^2 v(\xi)+\tilde{V}_-(\xi)v(\xi)={\cal E}_{\perp 0} v(\xi).
\label{onedimm}
\end{align}
\end{subequations}
Later the solutions of these equations will also be labeled with the value of energy ${\cal E}_{\perp 0}$, i.e. they will be denoted as $u_{{\cal E}_{\perp 0}}$ and $v_{{\cal E}_{\perp 0}}$. 
The first potential possesses bounds state for sure at least in the interesting range of parameters for which Figures~\ref{vv1} and~\ref{vv2} are sketched. Contrary, if one takes $\tilde{V}_-$ instead of $\tilde{V}_+$ no bound states exist. This can be easily seen by inspecting~(\ref{vt}) from which one can infer that $V_+(\xi)$ ($V_-(\xi)$) grows (declines) linearly as $\xi\rightarrow\infty$, as well as by watching the Figures~\ref{vv1} and~\ref{vv2}. 

However, bound states are not stationary states of the full Hamiltonian. There are two sources of possible tunneling, which are considered below. One results from the spatial size of the vortex (in perpendicular direction), which creates an opportunity for tunneling through the barrier. If the extension of the light beam is limited in the radial direction, $\tilde{V}_+(\xi)$ vanishes beyond certain value $\xi_w$ (where $\xi_w=k \rho_w$, the latter roughly corresponding to the waist of the beam) creating an ordinary potential barrier through which the tunneling is possible even in the zeroth order of the perturbation calculation (i.e. for $\tilde{W}=0$).  The probability of this process is negligible in comparison with the other as estimated in subsection~\ref{tunnelb}.

The second possibility is connected with the fact that the theory possesses two channels, and there is no binding in the second channel at all. Thus another way of tunneling emerges: that into the second channel. This phenomenon should be much more important. For such a process to occur it is necessary to turn on the off-diagonal terms of the Hamiltonian represented by $\tilde{W}$. This case is dealt with in subsection~\ref{tunnelc} in the perturbative manner. 

Let us turn below to the operator $\tilde{W}$ and verify, whether this kind of calculation will be justified. The quantity $\gamma/\alpha=\sqrt{\hbar\omega/Mc^2}$ is very small (see also Fig.~\ref{vv1} and~\ref{vv2}). Let us estimate below $\tilde{W}$ taking the above into account. From~(\ref{lambda}), (\ref{theta}), (\ref{funchi}) and (\ref{sigma}) for $\gamma/\alpha\ll 1$ we find:
\begin{subequations}\label{appr}
\begin{align}
\Theta&=-\frac{\beta}{2}+O(\frac{\gamma}{\alpha}),\label{thapp}\\
\Lambda&=\frac{1}{2}\sqrt{\beta^2+\alpha^2\xi^2}+O(\frac{\gamma}{\alpha}),\label{laapp}\\
\sigma_\pm&=\sqrt{2}\left[\sqrt{\beta^2+\alpha^2\xi^2}(\sqrt{\beta^2+\alpha^2\xi^2}\mp\beta)\right]^{-1/2}+O(\frac{\gamma}{\alpha}),\label{siapp}\\
\chi_\pm&=\pm \left[2\sqrt{\beta^2+\alpha^2\xi^2}(\sqrt{\beta^2+\alpha^2\xi^2}\mp\beta)\right]^{-1/2}\nonumber\\
&\times\left[\begin{array}{c}\sqrt{\beta^2+\alpha^2\xi^2}\mp\beta \\ \mp \alpha\xi\end{array}\right]+O(\frac{\gamma}{\alpha}),\label{chiapp}
\end{align}
\end{subequations}
Collecting all terms of the above approximations, after some laborious calculations omitted here, we get up to $(\gamma/\alpha)^2$:
\begin{eqnarray}
\tilde{W}=&&\!\!\!\! \frac{|\gamma|\beta}{2(\beta^2+\alpha^2\xi^2)}\left(\frac{\alpha^2\xi}{\beta^2+\alpha^2\xi^2}-\partial_\xi\right)\left[\begin{array}{cc} 0 & 1\\ -1 & 0\end{array}\right]\nonumber\\
=&&\!\!\!\! \tilde{W}_0(\xi)\left[\begin{array}{cc} 0 & 1\\ -1 & 0\end{array}\right]
,
\label{appw}
\end{eqnarray}
which can also be written in the explicitly hermitian form:
\begin{align}
\tilde{W}=-\frac{|\gamma|\beta}{2}\,\frac{1}{\sqrt{\beta^2+\alpha^2\xi^2}}\partial_\xi\,\frac{1}{\sqrt{\beta^2+\alpha^2\xi^2}}\left[\begin{array}{cc} 0 & 1\\ -1 & 0\end{array}\right].
\label{appw1}
\end{align}
Due to the coefficient $\gamma$ this quantity is small and may be treated as a perturbation. Since $\tilde{W}$ is purely off-diagonal, it does not contribute to the first-order perturbative correction to the bound state energy ${\cal E}_{\perp 0}$. Hence ${\cal E}_\perp-{\cal E}_{\perp 0}=O(\gamma^2)$ and this correction may be neglected within our present approach. The existence of bound states of the equation~(\ref{msch}) at least in the perturbative sense will be then indubitable if the rate of eventual tunneling is sufficiently small. 
 
\subsection{Tunneling through the barrier}
\label{tunnelb}

While considering the tunneling through the barrier, we cannot limit ourselves to the paraxial approximation where magnetic field given by the formula~(\ref{field}). Both the height and the thickness of the barrier are important and we are condemned to use the full form of magnetic field~(\ref{fieldfull}). The correction $\tilde{W}$ does not play any role here since it is off-diagonal, so that the tunneling probability can be estimated from Eq.~(\ref{onedimp}).

For small values of $k_\perp$ (but still admitting $k_\perp\rho \gtrsim 1$), we can neglect terms containing the coefficient $a_-$. Moreover it should be noted that since $|J_2(x)|<0.5$~\cite{spanier}, then for the values of parameters considered in this work the following estimation holds
\begin{eqnarray}
\left|\frac{4B_\perp \cos(\zeta_z-2\varphi)J_2(k_\perp\rho)}{B_z}\right|&\!\!<2\left|\frac{B_\perp}{B_z}\right|\nonumber\\
=\left|\frac{2\alpha}{\beta-\alpha/\gamma}\right|&\!\!\approx 2|\gamma|\ll 1,
\label{asB}
\end{eqnarray}
which means that the $z$ component of the magnetic field of the wave may be neglected as compared to the external field (which is strong), and we get

\begin{align}\label{fieldfulla}
{\bm B}({\bm r},t)=\left[\begin{array}{c}
-2B_\perp \sin(\zeta_z-\varphi)J_1(k_\perp\rho)\\
2B_\perp \cos(\zeta_z-\varphi)J_1(k_\perp\rho)\\
B_z \end{array}\right],
\end{align}

This form of magnetic field leads to the identical equation as~(\ref{msch}), but with the modified matrix potential
\begin{widetext}
\begin{align}
V(\xi)=\left[\begin{array}{cc}\displaystyle \frac{\gamma(\kappa_z+1/4)}{2\alpha}-\frac{\beta}{2}+\frac{\gamma}{2\alpha}\, \frac{(m+3/2)(m+1/2)}{\xi^2} &\displaystyle  -\alpha J_1(\kappa_z\xi) \\ \displaystyle -\alpha J_1(\kappa_z\xi) & \displaystyle \frac{\gamma(-\kappa_z+1/4)}{2\alpha}+\frac{\beta}{2}+\frac{\gamma}{2\alpha}\, \frac{(m+1/2)(m-1/2)}{\xi^2} \end{array}\right],
\label{vpotf}
\end{align}
\end{widetext}

Now all formulas of the Sec.~\ref{feq} remain true, if we redefine
\begin{align}
\Lambda(\xi)=\left[\Theta(\xi)^2+\alpha^2J_1(\kappa_z\xi)\right]^{1/2},
\label{lambdaf}
\end{align}
together with
\begin{align}
\chi_\pm(\xi)=\left[\begin{array}{c}\displaystyle \Theta(\xi)\pm \Lambda(\xi)\\ \displaystyle -\alpha J_1(\kappa_z\xi)  \end{array}\right]\sigma_\pm(\xi),
\label{funphif}
\end{align}

\begin{figure*}[t!]
\begin{center}
\includegraphics[width=0.8\textwidth,angle=0]{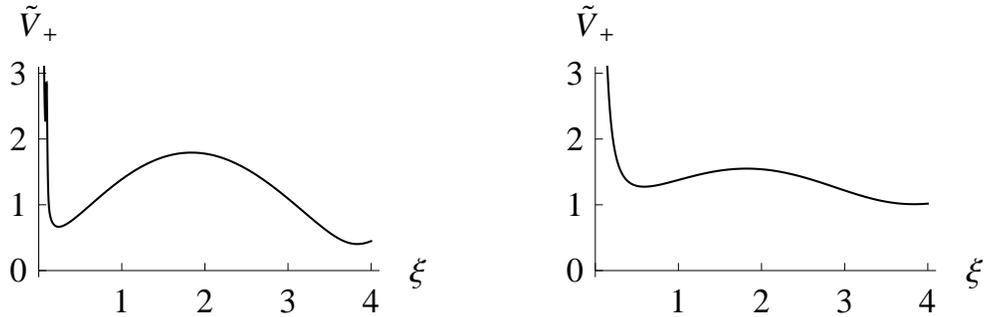}
\end{center}
\caption{Behavior of $\tilde{V}_+(\xi)$ derived from (\ref{vpotf}) for the values of parameters of Fig.~\ref{vv1} and \ref{vv2}.}
\label{vv3}
\end{figure*}

The behavior of the modified potential $\tilde{V}_+$ for both sets of data (those of Fig.~\ref{vv1} and \ref{vv2}) are drawn in Fig.~\ref{vv3}.
As it is well known both from the WKB method~\cite{wkb} and Millner's and Good's method~\cite{mg} the value of the tunneling probability is dictated by the typical exponential factor
\begin{equation}
\theta=e^{-2\int\limits_{x_0}^{x_0+d}\sqrt{\frac{2M}{\hbar^2}(V(x)-E)}\, dx},
\label{texp}
\end{equation}
where $d$ denotes the barrier thickness for a given energy. Since both the square root and $V(x)$ are concave functions in this region (see Fig.~\ref{vv3}), so it is obvious that
\begin{equation}
\theta<e^{-2\int\limits_{x_1}^{x_1+d}\frac{1}{2}\sqrt{\frac{2M}{\hbar^2}(V_{\mathrm max}-E)}\,dx}=e^{-d\sqrt{\frac{2M}{\hbar^2}(V_{\mathrm max}-E)}},
\label{test}
\end{equation} 
where `max' refers to the top of the barrier and $x_1$ denotes the right turning point for the classical motion. It can be rewritten as 
\begin{equation}
\theta<e^{-\sqrt{2}\xi_d\sqrt{\frac{\alpha}{\gamma}(\tilde{V}_{+{\mathrm max}}-{\cal E}_{\perp 0})}},
\label{thetasd}
\end{equation}
$\xi_d$ corresponding to $d$ in our dimensionless variables. The numerical prefactors of $\theta$ are marginal for the overall value of probability~\cite{raz}. 

For the first set of data the ground state energy can be estimated with the use of the uncertainty principle to be ${\cal E}_{\perp 0}\approx 0.67$. The maximal value of the potential is $\tilde{V}_{+{\mathrm max}}\approx 1.79$ and the barrier thickness $\xi_d\approx 3.12$. The barrier is then relatively high and thick which results in the extremely small value of the factor 
$$\theta < 6.7\times 10^{-36}.$$
The value of $\theta$ determines the transmission probability for each individual hitting the potential wall by the particle. To find the probability per unit time it should be multiplied by the number $n$ of hits per unit time. This can be estimated from the bound state energy as 
\begin{equation}
n\lesssim \frac{p_{\mathrm max}}{M|x_0-x_1|}\lesssim\frac{\sqrt{2M(E-V_{\mathrm min})}}{M|x_0-x_1|},
\label{nhi}
\end{equation}
where $x_0$ is the left turning point. For the dimensionless variables used throughout the work this reads:
\begin{equation}
n\lesssim\frac{\sqrt{2(\gamma/\alpha)^3({\cal E}_{\perp 0}-\tilde{V}_{+{\mathrm min}})}}{|\xi_0-\xi_1|}\, \omega\approx 3.6\cdot 10^{-4}\omega,
\label{nhim}
\end{equation}
leading to the negligible value
\begin{equation}
\frac{\Gamma}{\omega}\lesssim 2.4\cdot 10^{-39}.
\label{gg1}
\end{equation}

In the second case the potential cavity is shallower and the barrier thiner in a visible way. This is also reflected by the results of our previous work on classical motion~\cite{ibbtr2}, where the comparison of Figures~5 and~8 reveals much larger size of the trap in the latter case. For this data the trap is less efficient. In quantum theory this should entail the significant increase of the factor $\theta$. We find ${\cal E}_{\perp 0}\approx 1.28$ and $\tilde{V}_{+{\mathrm max}}\approx 1.55$. The barrier thickness gets reduced to $\xi_d\approx 2.21$. Consequently we can estimate~(\ref{thetasd}) to be about 
$$\theta < 8.4\times 10^{-8},$$
which is much larger but still very small. For $n$ we obtain
\begin{equation}
n\lesssim 7.9\cdot 10^{-4}\omega,
\label{nhim2}
\end{equation}
which gives
\begin{equation}
\frac{\Gamma}{\omega}\lesssim 6.6\cdot 10^{-11}.
\label{gg2}
\end{equation}

As we will see in the following subsection the direct tunneling through the barrier (i.e. within the first channel) may be practically neglected. A heavy particle with very short Compton wavelength cannot penetrate a barrier too deep without strongly violating the energy conservation. A much more essential effect is connected with the tunneling into the second channel at which the particle is kicked off out of the vortex field. 

\subsection{Tunneling to the other channel}
\label{tunnelc}

In order to consider the tunneling into the second channel we assume that beyond $\xi=\xi_w$ one has $\tilde{V}_-(\xi)=0$. Otherwise the potential of the scattering sector would be unphysical leading to the Hamiltonian unbounded below. This aspect was inessential for the evolution of solutions restricted to the first channel but must be taken into account in the case of inter-channel transitions. Consistently we modify the formula~(\ref{ev}) for $\tilde{V}_-$ by including the factor $\Theta(\xi_w-\xi)$ and henceforth this is the new meaning of that symbol. As to $\tilde{V}_+$ it is left unmodified consequently to the results of the previous subsection which clearly indicate that direct tunneling through the field barrier is improbable and from that point of view the binding potential may be treated as extending to infinity with no essential change of conclusions.

The only realistic tunneling process consists, therefore, on flipping the magnetic moment (keeping the constant of motion $H_2$ given by (\ref{cm2}) fixed), after which the particle gets ejected from the vortex field instead of being tunneled through the potential hump. 
Below we try to estimate the probability per unit time for this kind of a process. Aimed at simplifying the equation the independent variable $\xi$ (and similarly the parameter $\xi_w$) will now be rescaled as follows:
\begin{align}
\mathfrak{z}=\left|\frac{\alpha^{2}}{\gamma}\right|^{1/3}\xi.
\label{zx}
\end{align}
If this transformation is applied to the equation~(\ref{onedim}) it can be observed that the approximation of large $\mathfrak{z}$ is the same as that for $\gamma/\alpha\ll 1$. Therefore, its asymptotic form may be treated as applicable in the whole domain (possibly except a narrow -- and inessential -- interval close to the origin):
\begin{subequations}\label{onedimg}
\begin{align}
&-\partial_\mathfrak{z}^2 u(\mathfrak{z})+\mathfrak{z} u(\mathfrak{z})=\frac{2{\cal E}_{\perp 0}}{|\gamma\alpha|^{1/3}}\, u(\mathfrak{z}),
\label{onedimgp}\\
&-\partial_\mathfrak{z}^2 v(\mathfrak{z})-\mathfrak{z}\Theta(\mathfrak{z}_w-\mathfrak{z}) v(\mathfrak{z})=\frac{2{\cal E}_{\perp 0}}{|\gamma\alpha|^{1/3}}\, v(\mathfrak{z}).
\label{onedimgm}
\end{align}
\end{subequations} 
First let us deal with the former equation. Upon shifting the independent variable in order to cancel the right-hand side and denoting 
\begin{equation}
\mathfrak{z}_0=-2{\cal E}_{\perp 0}/(\gamma\alpha)^{1/3},
\label{z0}
\end{equation}
it becomes the ordinary Airy equation with solutions~\cite{spanier}:
\begin{align}
u(\mathfrak{z})=C_1\mathrm{Ai}(\mathfrak{z}+\mathfrak{z}_0)+C_2\mathrm{Bi}(\mathfrak{z}+\mathfrak{z}_0).
\label{onesolp}
\end{align}
For the bound state wave-function $u_{\mathfrak{z}_0}(\mathfrak{z})$ we require a sufficiently quick decline at infinity. Therefore, $C_2=0$ and $C_1$ may be denoted simply as $C$. On the other hand we must have $u_{\mathfrak{z}_0}(0)=0$, which means that $\mathfrak{z}_0$ is the first zero (as far as the ground state is considered) of the Airy function $\mathrm{Ai}$ (i.e $\mathfrak{z}_0\approx -2.338$). 

Using the integral~\cite{ov}:
\begin{eqnarray}
\int\limits_0^\infty &&\!\!\!\!\!\!{\mathrm Ai}^2(\mathfrak{z}+\mathfrak{z}_0)d\mathfrak{z}=\int\limits_0^\infty \partial_\mathfrak{z}(\mathfrak{z}+\mathfrak{z}_0){\mathrm Ai}^2(\mathfrak{z}+\mathfrak{z}_0)d\mathfrak{z}\nonumber\\
&& =- \mathfrak{z}_0{\mathrm Ai}^2(\mathfrak{z}_0)-2\int\limits_0^\infty (\mathfrak{z}+\mathfrak{z}_0){\mathrm Ai}(\mathfrak{z}+\mathfrak{z}_0){\mathrm Ai}'(\mathfrak{z}+\mathfrak{z}_0)d\mathfrak{z}\nonumber\\
&&=-\mathfrak{z}_0{\mathrm Ai}^2(\mathfrak{z}_0)-2\int\limits_0^\infty {\mathrm Ai}''(\mathfrak{z}+\mathfrak{z}_0){\mathrm Ai}'(\mathfrak{z}+\mathfrak{z}_0)d\mathfrak{z}\nonumber\\
&& =-\mathfrak{z}_0{\mathrm Ai}^2(\mathfrak{z}_0)-\int\limits_0^\infty \partial_\mathfrak{z}{{\mathrm Ai}'}^2(\mathfrak{z}+\mathfrak{z}_0)d\mathfrak{z}\nonumber\\
&& =-\mathfrak{z}_0{\mathrm Ai}^2(\mathfrak{z}_0)+{{\mathrm Ai}'}^2(\mathfrak{z}_0).\label{ai1}
\end{eqnarray}
the normalization constant $C$ can be found to be:
\begin{align}
C=[-\mathfrak{z}_0\mathrm{Ai}^2(\mathfrak{z}_0)+{\mathrm{Ai}'}^2(\mathfrak{z}_0)]^{-1/2}=\frac{1}{\mathrm{Ai}'(\mathfrak{z}_0)},
\label{cval}
\end{align}
where absolute value has been omitted as the derivative of the Airy function at $\mathfrak{z}=\mathfrak{z}_0$ is positive.

Now consider the function $v$ satisfying (\ref{onedimgm}). One can distinguish two characteristic regions: $0<\mathfrak{z}<\mathfrak{z}_w$ and $\mathfrak{z}>\mathfrak{z}_w$. The well-behaving and continuously differentiable wave-function $v(\mathfrak{z})$ corresponding to the quantum number $\mathfrak{z}_0$ can be written as:
\begin{align}
v_{\mathfrak{z}_0}(\mathfrak{z})=\left\{\begin{array}{lcl} G_{\mathfrak{z}_0}(\mathfrak{z}),&\mathrm{for} & 0<\mathfrak{z}<\mathfrak{z}_w,\\ 
\frac{1}{q_{\mathfrak{z}_0}}G'_{\mathfrak{z}_0}(\mathfrak{z}_w)\sin q_{\mathfrak{z}_0}(\mathfrak{z}-\mathfrak{z}_w)\\
\hspace{2ex}+ G_{\mathfrak{z}_0}(\mathfrak{z}_w)\cos q_{\mathfrak{z}_0}(\mathfrak{z}-\mathfrak{z}_w), &\mathrm{for}  & \mathfrak{z}>\mathfrak{z}_w,\end{array}\right.
\label{onesolm}
\end{align}
where $q_{\mathfrak{z}_0}=\sqrt{-\mathfrak{z}_0}$ and
\begin{align}
G_a(\mathfrak{z})=D_1\mathrm{Ai}(-\mathfrak{z}+a)+D_2\mathrm{Bi}(-\mathfrak{z}+a).
\label{defh}
\end{align}
To ensure the nonsingular behavior of $v_{\mathfrak{z}_0}(\mathfrak{z})$ at $\mathfrak{z}=0$ one has to put $D_1=D \mathrm{Bi}(a)$ and $D_2=-D\mathrm{Ai}(a)$. Above $\mathfrak{z}_w$, where the potential vanishes, we have ordinary trigonometric solutions.

The following integral
\begin{align}
\int\limits_0^\infty  v_a(\mathfrak{z})v_b(\mathfrak{z})d\mathfrak{z} = \pi \sqrt{-a}\left(G_a(\mathfrak{z}_w)^2+\frac{G'_a(\mathfrak{z}_w)^2}{q_a^2}\right)\delta(a-b).
\label{dval1}
\end{align}
can be used to fix $D$. The delta function on the r.h.s. comes from the integration between $\mathfrak{z}_w$ and infinity, where $v(\mathfrak{z})$ is given by the trigonometric functions. The off-diagonal (i.e. for $a\neq b$) integral can be shown to vanish if we use the following trick~\cite{ov}:
\begin{eqnarray}
&&\int\limits_0^{\mathfrak{z}_w} G_a(\mathfrak{z})G_b(\mathfrak{z})d\mathfrak{z} =\int\limits_0^{\mathfrak{z}_w}  \frac{G''_a(\mathfrak{z})}{a-\mathfrak{z}}\, \frac{G''_b(\mathfrak{z})}{b-\mathfrak{z}}d\mathfrak{z}\nonumber\\
&&=\frac{1}{b-a}\int\limits_0^{\mathfrak{z}_w} \left(\frac{G''_a(\mathfrak{z})}{a-\mathfrak{z}}\, G''_b(\mathfrak{z})-G''_a(\mathfrak{z})\frac{G''_b(\mathfrak{z})}{b-\mathfrak{z}}\right) d\mathfrak{z}\nonumber\\
&&=\frac{1}{b-a}\int\limits_0^{\mathfrak{z}_w} \left(G_a(\mathfrak{z}) G''_b(\mathfrak{z})-G''_a(\mathfrak{z})G_b(\mathfrak{z})\right) d\mathfrak{z}\label{f0i}\\
&&=\frac{1}{b-a}\int\limits_0^{\mathfrak{z}_w}\partial_\mathfrak{z} \left(G_a(\mathfrak{z}) G'_b(\mathfrak{z})-G'_a(\mathfrak{z})G_b(\mathfrak{z})\right) d\mathfrak{z}\nonumber\\
&&=\frac{1}{b-a}\left(G_a(\mathfrak{z}_w) G'_b(\mathfrak{z}_w)-G'_a(\mathfrak{z}_w)G_b(\mathfrak{z}_w)\right), \nonumber
\end{eqnarray}
and observe that this expression exactly cancels the terms coming from the integration of trigonometric functions in~(\ref{onesolm}) in the interval $(\mathfrak{z}_w, \infty)$.

On can show that the choice:
\begin{eqnarray}
D=&&\!\!\!\!\sqrt{\frac{2}{\pi}}\,\left(\frac{\alpha}{\gamma}\right)^{1/4}\label{dnorm}\\
&&\!\!\!\!\times\frac{|\mathrm{Bi}(\mathfrak{z}_0)|^{-1}}{\sqrt{\mathrm{Ai}(\mathfrak{z}_0-\mathfrak{z}_w)^2+(\alpha\gamma)^{1/3}/(2{\cal E}_{\perp 0})\, \mathrm{Ai}'(\mathfrak{z}_0-\mathfrak{z}_w)^2}}\,\nonumber
\end{eqnarray}
leads to the required normalization of the function $v(\xi)$:
\begin{align}
\int\limits_0^\infty v_{{\cal E}'_{\perp 0}}(\xi)v_{{\cal E}_{\perp 0}}(\xi) d\xi=\delta({\cal E}'_{\perp 0}-{\cal E}_{\perp 0}).
\label{normv}
\end{align}
Complex conjugations are omitted here and below since we are dealing with real functions only. 

The probability of tunneling per unit time is given by
\begin{align}
\Gamma=\frac{dP}{dt}=2\pi\omega\,\frac{\gamma}{\alpha}\,\sum\hspace{-3ex}\int\limits_i \delta({\cal E}_{\perp i}-{\cal E}_{\perp 0}) |I_{i0}|^2,
\label{Gamma}
\end{align}
where $i$ labels the continuum-spectrum states and $I_{i0}$ is the matrix element of the perturbation potential (the off-diagonal element of~(\ref{appw1})). According to what was said above, for the calculation of $I_{i0}$ within the present approximation the quantities $u_{\mathfrak{z}_0}$ and $v_{\mathfrak{z}_0}$ can be used as unperturbed wave-functions. Exploiting the delta function in~(\ref{Gamma}), we come to:
\begin{eqnarray}
\Gamma=&&\!\!\!\!\omega\frac{\pi}{2}\frac{\gamma}{\alpha}\left(\frac{\gamma}{\beta}\right)^2 \Bigg|\int\limits_0^\infty  \partial_\mathfrak{z}\left(\frac{u_{\mathfrak{z}_0}(\mathfrak{z})}{\sqrt{1+|\gamma \alpha|^{2/3}/\beta^2\; \mathfrak{z}^2}}\right)\nonumber\\
&&\times\frac{v_{\mathfrak{z}_0}(\mathfrak{z})}{\sqrt{1+|\gamma\alpha|^{2/3}/\beta^2\; \mathfrak{z}^2}}d\mathfrak{z}\Bigg|^2.
\label{Gamma1}
\end{eqnarray}
The coefficient $\frac{\pi}{2}\frac{\gamma}{\alpha}\left(\frac{\gamma}{\beta}\right)^2$ is about $10^{-6}$ for either sets of data. It should be noted that the strength of the inter-channel potential $\tilde{W}$, together the value of $\Gamma$, can be further reduced by the choice of $\beta$ to be very small. This point will be referred to at the end of this subsection.

The estimation of the integral can be done numerically assuming for instance the value of $\zeta_w\approx 4\pi|\alpha^2/\gamma|^{1/3}$, which roughly corresponds to two wavelengths. For the bound state energy ${\cal E}_{\perp 0}$ the value estimated in subsection~\ref{tunnelb} can be used and if instead the potential of Eq.~(\ref{onedimgp}) is taken for this assessment the obtained values turn out to be practically identical. It would be less accurate to use ${\cal E}_{\perp0}$ directly from Eq.~(\ref{z0}) since it would be underestimated due to the incorrect behavior of the approximated potential in the vicinity of $\zeta=0$ (the true potential does not vanish and has a positive minimum, which shifts the energy up). As it was mentioned, Eqs.~(\ref{onedim}) do not yield the correct form of the wave-functions close to the origin. We finally find
\begin{align}
\frac{\Gamma}{\omega}\approx 5.1\cdot 10^{-7},
\label{g1}
\end{align}
for the data of Fig.~\ref{vv1}, and 
\begin{align}
\frac{\Gamma}{\omega}\approx 7.2\cdot 10^{-6},
\label{g2}
\end{align}
for those of Fig.~\ref{vv2}. These values seem to be relatively large if time is measured in seconds or milliseconds. Nevertheless they prove that, with quantum effects involved, it is in principle possible to trap for a short time (microseconds) neutral particles through a very delicate mechanism relied on their magnetic moments interaction with magnetic field of the vortex. For instance for unstable particles the trapping time can be much larger than their lifetimes. 

The trapping time can be significantly prolonged, if the external magnetic field is well tuned so as to make the value of $\beta$ very small. For instance with the identical values of parameters as those of Fig.~\ref{vv1}, except for $\beta$, chosen now to be equal to $0.01$, one gets:
 
\begin{align}
\frac{\Gamma}{\omega}\approx 1.5\cdot 10^{-9}.
\label{g3}
\end{align}

This effect can be explained in classical terms as follows. The small value of the parameter $\beta$ corresponds to the tunning of $B_z$ so that the frequency of the Larmor precession of the magnetic moment around that field become close to the wave frequency. On could say that the external magnetic field keeps the magnetic moment synchronized in its rotation with the rotating vortex field. In these conditions the flipping of $\bm{\mu}$ necessary for the tunneling into the second channel, becomes less probable.

\begin{figure*}[t]
\begin{center}
\includegraphics[width=\textwidth,angle=0]{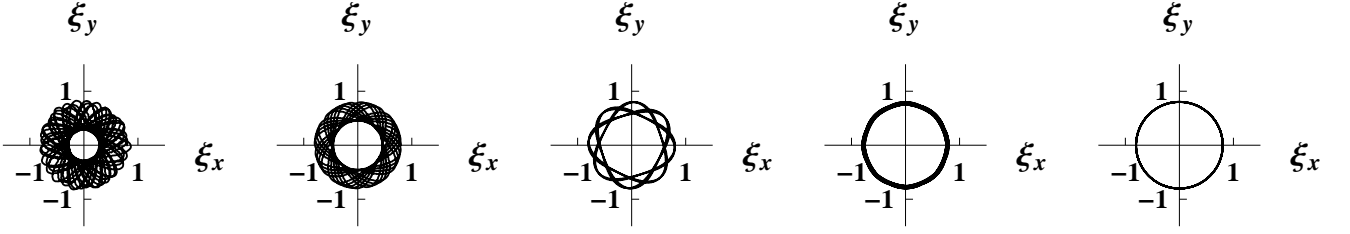}
\end{center}
\caption{Exemplary classical trajectory of a particle in a running wave considered in~\cite{ibbtr2} for subsequent decreasing values of parameter $\beta$:  $\beta=-0.03, -0.01, -0.005, -0.001, -0.0001$.}
\label{class}
\end{figure*}

This conclusion is supported by the observation referring to the classical motion of a particle for decreasing values of the parameter $\beta$. Upon precise tuning the initial particle state, there exist trajectories that become less and less chaotic as $\beta\rightarrow 0$, and turn into circles. As seen in Fig.~\ref{class} for extremely well adjusted value of the Larmor frequency to that of the vortex field or vice-versa, one can obtain a very stable (almost circular) trajectory, which suggests relatively strong binding of the particle by the vortex field.

\section{Summary}

The present work is concerned with the motion of a neutral quantum particle endowed with magnetic moment interacting with a certain special configuration of electromagnetic fields: a wave bearing orbital angular momentum (a vortex field) and a constant magnetic field aligned along the direction of propagation of the former (i.e. $z$ axis). It has been shown, by solving the appropriate Schr\"odinger equation, that this set up leads to the guidance of the particle along the vortex core and trapping it in the perpendicular direction. Due to the relatively weak strength of the interaction between the magnetic moment and the wave magnetic field the mechanism of capturing is delicate but still realizable.

The quantum theory confirms then the results obtained for the same values of parameters within classical mechanics. However, there appears purely quantum effect of tunneling which can play an important role for the particle being captured.

Two possible sources of tunneling were identified. The first one is connected with the fact, that the perpendicular size of the light beam is limited, thus creating a kind of a potential hump binding the particle. On the exterior of it the potential vanishes and a free motion is possible. The tunneling probability through this barrier was estimated to be extremely small and practically negligible in comparison with the second possibility.
The latter is connected with the spontaneous process of flipping the direction of magnetic moment to the opposite one. In these conditions the vortex field no longer keeps the particle bound but ejects it out of it. This kind of tunneling narrows the trapping time to microseconds. However this short time can be prolonged if one accurately tunes the external magnetic field. The idea is to make the Larmor frequency of precession of the magnetic moment around the $z$ axis close to that of the rotating magnetic field of the vortex. These circumstances make the flipping of spin much less probable since it is well synchronized with the rotating field. In consequence the trapping time can be lengthened
by a couple of orders of magnitude.

The same effect may be identified in classical mechanics. Naturally there is no tunneling there, but what is visible, is the stabilization of the particle orbit while approaching the resonance. The trajectory becomes much less chaotic and more circular, which corresponds to reducing the tunneling probability in the quantum case.

\section*{Acknowledgments}
I would like to thank Professor Iwo Bia\l ynicki-Birula for the inspiration and for many long discussions and valuable suggestions. The work was supported in part by Polish National Science Center Grant No. 2012/07/B/ST1/03347.

\appendix 
\section{Schr\"odinger-Pauli equation for the center-of-mass motion of the hydrogen atom}
\label{sc}

Let us start with the classical Hamiltonian of the electron-proton system in the external electromagnetic field:
\begin{eqnarray}
H=&&\!\!\!\!\frac{1}{2m_e}[{\bm p}_e-e{\bm A}({\bm r}_e,t)]^2+\frac{1}{2m_p}[{\bm p}_p-e{\bm A}({\bm r}_p,t)]^2\nonumber\\
&&\!\!\!\!-\frac{e^2}{4\pi\varepsilon_0|{\bm r}_e-{\bm r}_p|}-{\boldsymbol\mu}_e{\bm B},
\label{hamilta}
\end{eqnarray}
where indices $e$ and $p$ refer to electron and proton respectively. The magnetic moment of the proton has been neglected due to the very small ratio $m_e/m_p$.

We now follow the procedure elaborated in~\cite{schleich}. Introducing relative and center-of-mass coordinates:
\begin{subequations}
\label{newcoord}
\begin{align}
&{\bm r}={\bm r}_e-{\bm r}_p,\\
&{\bm R}=\frac{m_e{\bm r}_e+m_p{\bm r}_p}{m_e+m_p},
\end{align}
\end{subequations}
and momenta:
\begin{subequations}
\label{newmom}
\begin{align}
&{\bm p}=\frac{m_p{\bm p}_e-m_e{\bm p}_p}{m_e+m_p},\\
&{\bm R}={\bm p}_e+{\bm p}_p,
\end{align}
\end{subequations}
one can rewrite the Hamiltonian in the following form:
\begin{align}
H=&\frac{1}{2M}[{\bm P}-e({\bm r}{\boldsymbol\nabla}_{\bm R}){\bm A}({\bm R},t)]^2+\frac{1}{2 M_r}[{\bm p}-e{\bm A}({\bm R},t)]^2 \nonumber\\
&-e\,\frac{\Delta M}{M}({\bm r}{\boldsymbol\nabla}_{\bm R}){\bm A}({\bm R},t)]^2-\frac{e^2}{4\pi\varepsilon_0r}-{\boldsymbol\mu}_e{\bm B},
\label{hamilt1}
\end{align}
where $M=m_e+m_p$ is the total mass of the system, $M_r=m_em_p/(m_e+m_p)$ is the reduced mass and $\Delta M=m_p-m_e$.

While passing from~(\ref{hamilta}) to~(\ref{hamilt1}) we assumed, that the external field varies slowly on the scale imposed by the size of the hydrogen atom. In the present work we deal with wavelengths, which are $3$ or $4$ orders of magnitude larger than this size, so the above approximation is well justified. Therefore, one can write:
\begin{subequations}
\label{apa}
\begin{align}
{\bm A}({\bm r}_e,t)=&\,{\bm A}({\bm R}+\frac{m_p}{m_e+m_p}{\bm r},t)\nonumber \\ \simeq &\,{\bm A}({\bm R},t)+\frac{m_p}{m_e+m_p}({\bm r}{\boldsymbol\nabla}_{\bm R}){\bm A}({\bm R},t),\\
{\bm A}({\bm r}_p,t)=&\,{\bm A}({\bm R}-\frac{m_e}{m_e+m_p}{\bm r},t)\nonumber\\ \simeq &\,{\bm A}({\bm R},t)-\frac{m_e}{m_e+m_p}({\bm r}{\boldsymbol\nabla}_{\bm R}){\bm A}({\bm R},t).
\end{align}
\end{subequations}

In order to express the Hamiltonian in terms of physical fields $\bm E$ and $\bm B$ only, we first perform the Legendre transformation on~(\ref{hamilt1}) and find the appropriate Lagrangian:
\begin{align}
L=&\frac{1}{2}\,M\dot{\bm R}^2+\frac{1}{2}\,M_r\dot{\bm r}^2+e\dot{\bm r}{\bm A}+\frac{e^2}{4\pi\varepsilon_0r}+{\boldsymbol\mu}_e{\bm B}\nonumber\\
&+e\dot{\bm R}({\bm r}{\boldsymbol\nabla}_{\bm R}){\bm A}+e\,\frac{\Delta M}{M}\dot{\bm r}({\bm r}{\boldsymbol\nabla}_{\bm R}){\bm A}.
\label{lagr}
\end{align}
This Lagrangian may be modified by subtracting from it a total derivative over time:
\begin{align}
\tilde{L}=L-\frac{d}{dt}\left[e{\bm r}{\bm A}+\frac{e}{2}\,\frac{\Delta M}{M}\dot{\bm r}({\bm r}{\boldsymbol\nabla}_{\bm R}){\bm A}\right].
\label{sub}
\end{align}
In the radiation gauge, where $A_0=0$ and ${\boldsymbol\nabla}{\bm A}=0$, we have ${\bm E}=-\dot{\bm A}$, and~(\ref{sub}) may be given the form:
\begin{align}
\tilde{L}&=\frac{1}{2}\,M\dot{\bm R}^2+\frac{1}{2}\, M_r\dot{\bm r}^2+\frac{e^2}{4\pi\varepsilon_0r}+{\boldsymbol\mu}_e{\bm B}+e{\bm r}{\bm E}\label{lnew}\\
&-e\dot{\bm R}({\bm r}\times {\bm B})-\frac{e}{2}\,\frac{\Delta M}{M}\dot{\bm r}({\bm r}\times {\bm B})+\frac{e}{2}\,\frac{\Delta M}{M}({\bm r}{\boldsymbol\nabla}_{\bm R})({\bm r}{\bm E}),
\nonumber
\end{align}
where we made use of the identity:
\begin{align}
{\boldsymbol\nabla}_{\bm R}({\bm r}{\bm A})-({\bm r}{\boldsymbol\nabla}_{\bm R}){\bm A}={\bm r}\times{\bm B}.
\label{ide}
\end{align}

Now, performing the inverse Legendre transformation, we obtain the modified Hamiltonian (from now on we omit `tilde'):
\begin{align}
H=&\frac{1}{2M}\left({\bm P}+e{\bm r}\times{\bm B}\right)^2+\frac{1}{2 M_r}\left({\bm p}+\frac{e}{2}\,\frac{\Delta M}{M}\,{\bm r}\times{\bm B}\right)^2 \nonumber\\
&-\frac{e^2}{4\pi\varepsilon_0r}-{\boldsymbol\mu}_e{\bm B}-e{\bm r}{\bm E}-\frac{e}{2}\,\frac{\Delta M}{M}({\bm r}{\boldsymbol\nabla}_{\bm R})({\bm r}{\bm E}).
\label{hnew}
\end{align}

We are not interested in the internal atomic structure but in the motion of a neutral particle, as for instance a hydrogen atom, as a whole. We, therefore, separate the center-of-mass motion from the internal one. All terms that do not depend on $\bm R$ are then irrelevant. Furthermore, since $r\sim a_0$, where $a_0$ is the Bohr radius, we can neglect terms, which are of order of $r^2$. The same refers to those, that are linear in $r$ but are accompanied by the magnetic field $\bm B$. The term $e{\bm r}{\bm E}$ is of order of $\varepsilon_0 a_0^3 E^2$ and may be omitted too. One should also remember, that the external electromagnetic fields ${\bm E}$ and ${\bm B}$, due to the approximations~(\ref{apa}), depend only on $\bm R$. We are then left with a rather obvious form:
\begin{align}
H_{CM}=\frac{{\bm P}^2}{2M}-{\boldsymbol\mu}_e{\bm B},
\label{hcm}
\end{align}
which will constitute the starting point for considering the Schr\"odinger-Pauli equation for the hydrogen atom, as well as for any neutral particle endowed with the magnetic moment as neutron for instance.


\begin{thebibliography}{99}
\bibitem{schi} M. Schiffer {\em et al.}, Appl. Phys. {\bf B 67}, 705(1998).
\bibitem{kupp} S. Kuppens {\em et al.}, Phys. Rev. {\bf A 58}, 3068(1998).
\bibitem{rho1} D.P. Rhodes, G.P.T. Lancaster and J.G. Livesey, Opt. Commun. {\bf 214}, 219(2002).
\bibitem{glo} D. McGloin {\em et al.}, Opt. Expr. {\bf 11}, 158 (2003).
\bibitem{rho2} D.P.Rhodes {\em et al.}, J. Mod. Opt. {\bf 53}, 547 (2006).
\bibitem{shan} M.A. Ol’Shanii, Yu.B. Ovchinnikov and V. S. Letokhov,
Opt. Commun. {\bf 98}, 77(1993).
\bibitem{mark} S. Marksteiner{\em et al.}, Phys. Rev. {\bf A 50}, 2680(1994).
\bibitem{renn1} M.J. Renn {\em et al.}, Phys. Rev. Lett. {\bf 75}, 3252(1995).
\bibitem{renn2} M.J. Renn {\em et al.}, Phys. Rev. {\bf A 53}, R648(1996).
\bibitem{yin} J. Yin {\em et al.}, J. Opt. Soc. Am. {\bf B 15}, 25(1998).
\bibitem{arlt1} J. Arldt, T. Hitomi and K. Dholakia, Appl. Phys. {\bf B 71}, 549 (2000).
\bibitem{arlt2} J. Arldt, {\em et al.}, Phys. Rev. {\bf A 63}, 063602(2001).
\bibitem{wang} Z. Wang, Y. Dong and Q. Lin, J. Opt. {\bf A 7}, 147(2005).
\bibitem{mielnik} B. Mielnik and D.J. Fern\'andez, J. Math. Phys. {\bf 30}, 537(1989).
\bibitem{fer} D.J. Fern\'andez and L.M. Nieto, Phys. Lett. {\bf A 161}, 202(1991).
\bibitem{QL} Qiong-gui Lin, Phys. Rev. {\bf A 63}, 012108(2001).
\bibitem{shapiro} V. E. Shapiro, Phys. rev. {\bf A 54}, R1018(1996).
\bibitem{ibbtr2} I. Bia{\l}ynicki-Birula and T. Rado\.zycki, Phys. Rev. {\bf A 93}, 063402(2016).
\bibitem{trojan} I. Bialynicki-Birula, Z. Bialynicka-Birula and B. Chmura, Laser Phys. {\bf 15}, 1371(2005).
\bibitem{ibbprl} I. Bialynicki-Birula, Phys. Rev. Lett. {\bf 93}, 020402(2005).
\bibitem{ibbtr1} I. Bia{\l}ynicki-Birula and T. Rado\.zycki, Phys. Rev. {\bf A 73}, 052114(2006).
\bibitem{spanier} J. Spanier and K.B. Oldham, {\em An atlas of functions}, Springer, 1987.
\bibitem{wkb} P.M. Morse and H. Feshbach, {\em Methods of Theoretical Physics}, McGraw-
Hill, New York 1953.
\bibitem{mg} S.C. Miller, Jr. and R.M. Good, Jr., Phys. Rev. {\bf 91}, 174(1953).
\bibitem{raz} M. Razavy, {\em Quantum Theory of Tunneling}, World Scientific, Singapore 2014.
\bibitem{ov} O. Val\'ee and M. Soares, {\em Airy Functions and Applications to Physics}, World Scientific, London 2004.
\bibitem{schleich} W.P. Schleich, {\em Quantum Optics in Phase Space}, Wiley, Berlin 2001.
\end{thebibliography}
\end{document}